%% file: rfid_ieee.tex
\documentclass[letterpaper,10 pt,conference]{IEEEtran}
\usepackage{amsmath}
\usepackage[dvips]{graphicx}
\usepackage{epic,eepic}
\usepackage{theorem}

\newtheorem{thm}{Theorem}
\newtheorem{lem}[thm]{Lemma}
\newcommand{\veck}{\ensuremath{\bar{k}}}

\newcommand{\eqnref}[1]{Eqn.~(\ref{#1})}
\newcommand{\figref}[1]{Fig.~\ref{#1}}

\begin{document}
\title{Reliable Identification of RFID Tags Using Multiple Independent Reader Sessions}

\author{\IEEEauthorblockN{Rasmus Jacobsen, Karsten Fyhn Nielsen, Petar Popovski and Torben Larsen}
\IEEEauthorblockA{Department of Electronic Systems, Aalborg University, Denmark \\
\{raller, kfyhn, petarp, tl\}@es.aau.dk}}

\maketitle

\begin{abstract}

Radio Frequency Identification (RFID) systems are gaining momentum in various applications of logistics,
inventory, etc.
A generic problem in such systems is to ensure that the RFID readers can reliably read a set of RFID tags,
such that the probability of missing tags stays below an acceptable value.
A tag may be missing (left unread) due to errors in the communication link towards the reader e.g. due to obstacles in the radio path.
The present paper proposes techniques that use multiple reader sessions,
during which the system of readers obtains a running estimate of the probability to have at least one tag missing.
Based on such an estimate, it is decided whether an additional reader session is required.
Two methods are proposed, they rely on the statistical independence of the tag reading errors across different reader sessions,
which is a plausible assumption when e.g. each reader session is executed on different readers.
The first method uses statistical relationships that are valid when the reader sessions are independent.
The second method is obtained by modifying an existing capture--recapture estimator.
The results show that, when the reader sessions are independent,
the proposed mechanisms provide a good approximation to the probability of missing tags,
such that the number of reader sessions made, meets the target specification.
If the assumption of independence is violated, the estimators are still useful,
but they should be corrected by a margin of additional reader sessions to ensure that the target probability of missing tags is met.
\end{abstract}

\begin{IEEEkeywords}
Missing tag problem, set cardinality estimation, error probability estimation, RFID networks
\end{IEEEkeywords}

\section{Introduction}
\label{intro}
\label{introduction}
RFID technology features a growing set of applications for
identification of various objects. The applications span from
simply identifying objects, serving as more informative barcodes,
gathering of sensory data and holding private/confidential
information \cite{supplychain}\cite{passports}\cite{library}.
The advantages of RFID technology include
the low cost per tag and the low energy consumption, which lets
them have a very long lifetime \cite{finkenzeller}. The passive RFID tags
represent a category of tags that does not have power supply, they
rely solely on the signal sent from a reader to power their
circuitry, and to respond by backscattering the signal \cite{finkenzeller}.

The communication paradigm in the passive RFID systems is based on
request/response: In the first step, the reader sends an
interrogation signal to the tags within its range. In the second
step the tags send their response to the reader by backscattering the signal.
If multiple tags simultaneously reply to the reader,
then the reader experiences tag collision. Hence, the reader
should run a certain anti-collision protocol (also called
collision-resolution or arbitration protocol) in order to
successfully resolve each tag in its proximity. There are
various anti-collision protocols,
which are in general divided into two groups, ALOHA--based \cite{epc}\cite{Vogt02efficientobject} and
tree-based \cite{708695}\cite{conflict_multiplicity}.

Regardless of the actual arbitration protocol used, after a single
run of the protocol is terminated, the reader has the identities
of the tags in its proximity. In the ideal case, when there are no
transmission errors and the only error experienced at the reader
is due to the tag collisions, then one can be certain that all the
tag identities have been collected during the arbitration process.
However, errors do occur if either the query from a reader is not
received correctly at a tag or the tag reply is not received at
the reader. In principle, if a tag is at a blind
spot~\cite{tagmark}, then the communication between the tag and
the reader is always in error. The probability that a tag is at a
blind spot can be substantial and is primarily determined by the
physical disposition of the tag, but also by the material to which
the tags are affixed. In \cite{gs1} it is shown, that if a tag is
attached to solar cream, the probability of not resolving a tag is
30\% and with mineral water it is 67\%. The error probability can
vary a lot, increasing the probability of missing one or more tags
completely. In summary, if during the arbitration protocol the
link between a reader and a tag is in error, then this tag is
not identified at the end of the protocol run. This is defined
as \textit{the missing tag problem}.

There are multiple approaches to minimize the probability of
missing a tag. In \cite{tag_id_patent}, a method for determining
group completeness in an RFID network is described, based on each
tag storing one or more references to surrounding tags. The
resolved tags and the references are compared, and if not all
references are resolved, the reading/comparison is repeated.
Thereby the reader knows with high probability if tags are
missing. This method is targeting rather static constellations of
tags, e.g. goods on pallets. Another approach is presented in
\cite{tagmark}, where a method for resolving a set of RFID tags is
presented. This is done by using two independent samples, in this
case a database and RFID readings. These two samples are used as
in a classical capture--recapture model \cite{citeulike:781430} to derive
estimators for the tag set cardinality.

The paper is organized as follows.
An overview of the problem and an intuitive explanation is presented in the next section in the case of two reader sessions, followed by the system model in Section \ref{system_model}.
Derivation of estimators for two reader sessions is in Section \ref{proposed_solution}.
The estimators are generalized to multiple reader sessions in Section \ref{framework},
and the estimators in the case of two reader sessions are evaluated analytically in Section \ref{analytical_analysis}.
In Section \ref{experimental_analysis},
simulations show the performance of the proposed estimators in scenarios with both dependent and independent reader sessions.
The work is concluded in Section \ref{conclusion}.

\section{Problem Definition}

The main idea in this paper is to use several independent readings
of the tag set that consists of $N$ tags. One reading of the tag
set consists of one run of the arbitration protocol, and is denoted a reader session. In each
reader session the probability that a given tag is not read is $p$.
Reader sessions are independent, when the probability that each
tag is read in one reader session is independent of it being read
in another. The value of the error probability $p$ and tag
cardinality $N$ are not known a priori. At this point it is
natural to ask: How can we assure, or at least attempt, to make
the readings independent? Here are two plausible examples:
\begin{enumerate}
    \item Before the next reader session with the same reader,
the tagged items are physically displaced/shuffled and it can be
assumed that such an action ``resets'' the physical links and
generates error with probability $p$.
    \item If one reader with multiple antennas or multiple readers
    are located at different positions, but remain in communication range with the same tags,
    the reader sessions may be assumed independent.
\end{enumerate}
A scenario that encompasses both cases is the one with a conveyor
belt, along which several readers are deployed. It should be noted
that these are ways to aim for independence and simulations show
how the methods introduced underperform when the independence assumption
does not hold.

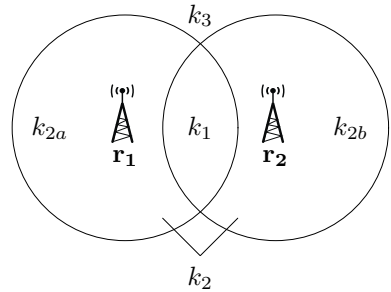
\begin{figure}
\centering
\input{venn}
\caption{Venn diagram of the possible tag sets for two reader sessions $r_1$ and $r_2$.
$k_1$ is the number of tags found in common in both reader sessions,
$k_{2a}$ is the number of tags only found in reader session $r_1$, $k_{2b}$ only in reader session $r_2$.
The set $k_2$ is given by the sum of $k_{2a}$ and $k_{2b}$.
An unobserved number of tags, $k_3$, may exist.}
\label{fig:venn}
\end{figure}

The basic idea of our approach leverages on the recent ideas about
cooperative readers~\cite{tree_based} that can jointly infer
statistical information about the set of tags $S$ in range. In order
to illustrate the idea, consider the case with two readers each having a reader session, $r_1$ and $r_2$ respectively.
The probability that a tag
is not read in reader session $r_i$ is $p$. After the two reader
sessions are terminated, the readers exchange information about
the tag identities they have gathered.
Let $k_1$ denote the subset of tags that have been read in both reader session $r_1$ and $r_2$.
Let $k_{2a}(k_{2b})$ be the subset of tags that are read only in $r_1 (r_2)$.
This is schematically represented in \figref{fig:venn}.
There is also a set of $k_3$ tags that are not read in either of the reader sessions.
Let $\hat{p}$ and $\hat{N}$ denote the estimates of $p$ and $N$, respectively.
Based on the expected values for $k_1$, $k_{2a}$, and $k_{2b}$, one can write:
    \begin{eqnarray}\label{eq:intro_example}
       k_1 &=& \hat{N} (1-\hat{p})^2 \nonumber \\
       k_{2a}+k_{2b} &=& 2\hat{N}  (1-\hat{p})\hat{p}
     \end{eqnarray}
Using these two equations,
we can obtain values for $\hat{p}$ and $\hat{N}$.
Based on that, we can estimate the expected value of the number of missing tags $k_3$.
Furthermore, we can estimate the probability of having at least one tag missing and,
if this probability is above a threshold value,
we can perform additional readings.
This process is generalized by devising methods to obtain $\hat{p}$ and $\hat{N}$ from three or more independent readings.
The objective is to create a sequential decision process in which,
after the $R$th reader session (arbitration protocol run),
we calculate the probability of having a tag remaining and,
if this probability is above a threshold value,
we carry out the $(R+1)$th reader session.

For the general case of $R>2$ reader sessions,
we propose two classes of estimators.
One class of estimators is emerging from the generalization of \eqnref{eq:intro_example}.
The other class of estimators is obtained by extending a classical capture--recapture result by Schnabel~\cite{1938},
in order to be able to estimate the error probability $p$.
These estimator classes are the major contributions of this paper,
along with the overall idea of sequential decision process in dealing with the missing tag problem.

\section{System Model}
\label{system_model}
\label{system_model} The system considered consists of multiple
readers and tags. Each reader can have multiple reader
sessions, defined as a session in which the reader runs its
arbitration protocol, trying to resolve the entire tag set. The
outcome of a reader session is a set specifying the tags resolved
by a reader in a reader session. The sets are assumed to contain no
errors, meaning that if a tag successfully backscatters a signal to the reader
without collision, then the tag is present in a set and the tag \emph{is}
resolved. The reader sessions are assumed to be coordinated in a
way that the readers do not interfere with each other i.e. the
reader collision problem \cite{reader_collision_problem} does not occur.

Throughout the paper we assume independent reader sessions,
except in Section \ref{dependent_reader_sessions}, where we introduce the correlated model
for evaluation of dependent reader sessions. In a given session,
each tag will, with probability $p$, be in a blind spot, i.e. not
being able to communicate with the reader. The complete tag set
$S$ contains $N$ tags and remains unchanged through the reader sessions.
The probability of error (blind spot) $p$ is identical for each
tag in each reader session. That is, for a given reader session
and a given tag, the tag is made unreadable with probability $p$, independently of the
other tags and previous reader sessions.

We assume independence \emph{across the tags}: the
event whether a tag $\tau_m$ is readable does not depend on the
event whether another tag $\tau_l$ is readable. On the other hand,
we introduce correlation by defining conditional
probabilities that the tag $\tau_m$ is readable in reader session $r_{i+1}$ provided that the same tag $\tau_m$ was readable/not
readable in $r_i$. The conditional probabilities in
the correlated case is selected such that the expected number
of non--readable tags in the reader session $r_{i+1}$ remains $Np$.
This is physically plausible, as we should not be able to improve
the overall readability of the tags set through a random physical
displacement.

\section{Proposed Solution}
\label{proposed_solution}

Four random variables, $K_1$, $K_{2a}$, $K_{2b}$ and $K_3$,
follow the multinomial distribution,
and describes the number of tags in the sets $k_1$, $k_{2a}$, $k_{2b}$, and $k_3$, respectively,
see \figref{fig:venn}.
The probability of a tag being read in the first reader session is $(1-p)$,
and in two reader sessions $(1-p)^2$.
The probability of a tag being read in the first but not in the next (and vice versa) is $(1-p)p$,
and not read at all is $p^2$.
This gives the probability mass function (pmf)
\begin{align*}
&\Pr[K_1=k_1,K_{2a}=k_{2a},K_{2b}=k_{2b},K_3=k_3]\\
&\quad={N \choose k_1,k_{2a},k_{2b},k_3}(1-p)^{2k_1}((1-p)p)^{k_{2a}+k_{2b}}p^{2k_3}.
\end{align*}
Lets define one more random variable, $K_2$, being the sum of $K_{2a}$ and $K_{2b}$,
then, assuming that they are independent gives the pmf
\begin{align*}
&\Pr[K_1=k_1,K_2=k_2,K_3=k_3]\\
&\quad={N \choose k_1,k_2,k_3}(1-p)^{2k_1}(2(1-p)p)^{k_2}p^{2k_3}.
\end{align*}
The expected values of the random variables are
\begin{align*}
E[K_1]&=N(1-p)^2\\
E[K_2]&=E[K_{2a}]+E[K_{2b}]=2N(1-p)p\\
E[K_3]&=Np^2
\end{align*}
When measured values of $k_1$ and $k_2$ are found,
and by assuming that they are close to their respective expected value,
we assume the following approximation
\begin{align}
\nonumber k_1&=\hat{N}(1-\hat{p})^2&\approx E[K_1]\\
\nonumber k_2&=2\hat{N}(1-\hat{p})\hat{p}&\approx E[K_2]\\
k_3&=\hat{N}\hat{p}^2&\approx E[K_3]
\label{eqkett}
\end{align}
Based on this, an estimate of $\hat{p}$ can be found,
by taking a ratio based on the set relationship, namely
\begin{equation}
\frac{k_1}{k_2}=\frac{\hat{N}(1-\hat{p})^2}{2\hat{N}(1-\hat{p})\hat{p}}\Rightarrow\hat{p}=\frac{k_2}{2k_1+k_2}.
\label{ptworeaders}
\end{equation}
Note that the (unknown) tag set cardinality $\hat{N}$ is cancelled out.
Using this estimator an estimate of $N$ can be found,
based on the fact that $E[K_3]=Np^2\approx \hat{k}_3=\hat{N}\hat{p}^2$ and $\hat{N}=k_1+k_2+\hat{k}_3$.
This yields an estimate of $N$ for two reader sessions
\begin{equation}
k_1+k_2=\hat{N}-\hat{k}_3=\hat{N}(1-\hat{p}^2)\Leftrightarrow\hat{N}=\frac{k_1+k_2}{1-\hat{p}^2},
\label{ntwocase}
\end{equation}
where $\hat{p}$ is given by $k_1$ and $k_2$ in \eqnref{ptworeaders}.

When estimates of $p$ and $N$ have been obtained, the probability of missing one or more tags can be calculated.
As the probability of missing one tag in one reader session is $\hat{p}$,
the probability of not missing $\hat{N}$ tags in two reader sessions is $(1-\hat{p}^2)^{\hat{N}}$.
This gives the estimate of the probability of missing at least one tag as
\[\hat{p}_M = 1-(1-\hat{p}^2)^{\hat{N}}.\]
If this probability is large, it is likely that tags are left unread.
It is possible to improve the estimates by making more than two reader sessions.
This is described in the following section.

\section{Generalization to Multiple Reader Sessions}
\label{framework}
To provide better estimates, the two-reader session case is extended to support more independent reader sessions.
The observable sets $k_1$ and $k_2$ are extended by defining a vector, $\veck=[k_1,\ldots,k_R]^T$,
which holds information about how many tags were found in how many reader sessions.
The first entry specify the number of tags found in $R$ reader sessions,
the second the number of tags found in $R-1$ reader sessions (regardless of which reader sessions) and so on.
Each element in $\veck$ is defined by extending \eqnref{eqkett} to:
\begin{align}
k_i=\hat{N} {R\choose R-(i-1)} (1-\hat{p})^{R-(i-1)} \hat{p}^{i-1}\approx E[K_i],
\label{eqn:k_i}
\end{align}
where $i = \{1,2,...,R\}$.
However, when extending to more than two reader sessions,
there are more relationships between the sets.
In the two-reader session case the measurable sets are $k_1$ and $k_2$, and the ratio $k_1/k_2$ is used (see \eqnref{ptworeaders}),
but others exist, namely $k_1/(k_1+k_2)$ and $k_2/(k_1+k_2)$ which lead to the same estimator for $p$.
When the number of reader sessions increases, then the number of measurable sets and the number of possible ratios increases,
e.g. for three reader sessions, the sets are $k_1$, $k_2$ and $k_3$, and possible ratios are $k_1/k_2$, $k_1/k_3$, $k_2/k_3$, $k_1/(k_2+k_3)$, $k_2/(k_1+k_2)$, etc.
Therefore we do not have one good ratio with equally weighted sets,
and common to almost all of the ratios is that an explicit expression for $\hat{p}$ does not exist,
and $\hat{p}$ needs therefore to be calculated numerically.
Before explaining some of the possible estimators of $p$,
the estimator of $N$ and the method of calculating the probability of missing one or more tags are explained.

The estimator of $N$ for $R>2$ reader sessions is based on the estimator of the tag set cardinality from \cite{conflict_multiplicity}.
The percentage of resolved tags is $(1-p^R)$, and the number of resolved tags is the sum of $k_i$s,
therefore the tag set cardinality can be generalized to
\[
	\hat{N}=\frac{\sum_{i=1}^R k_i}{1-\hat{p}^R}
\]

The estimate of the probability of missing at least one tag is extended to:
\begin{align}
\hat{p}_M = 1-(1-\hat{p}^R)^{\hat{N}}.
\label{eqn:p_m}
\end{align}
This estimator is useful if an application requires that the probability of one or more missing tags shall be lower than some threshold, $t_1$ (e.g. $t_1=10^{-5}$).
If $\hat{p}_M>t_1$ another reader session is required.
A new $\hat{p}_M$ is estimated for each reader session until the threshold set by the application is satisfied.

As this estimated error, $\hat{p}_M$,
is based on estimates of $p$ and $N$,
it relies on these being ``good''.
Therefore it can be necessary to either 1) add an artificial bias to $\hat{p}_M$,
or 2) perform an extra reading after the criteria is satisfied.
This is because $\hat{p}_M$ could be lower than $t_1$ in some cases where it should not,
as $\hat{p}$ could be underestimated.

Where both the estimates of $N$ and $p_M$ are straightforward to compute given $\hat{p}$,
$\hat{p}$ itself is not easy to compute directly,
because the estimate is found based on a ratio of sums of elements in $\veck$,
and the performance of the estimator depends on choosing a good ratio.

\section{Error Probability Estimators}
An estimator of the error probability is defined by which elements from $\veck$ are included in the ratio's numerator and denominator respectively.
Two \emph{window functions}, $\phi_n(k_i)$ and $\phi_d(k_i)$, are used to describe which elements are included.
The ratio is then defined as:
\begin{equation}
\frac{\sum_{i=1}^R\phi_n(i)k_i}{\sum_{i=1}^R\phi_d(i)k_i}
=\frac{\sum_{i=1}^R\phi_n(i){R\choose R-(i-1)} (1-\hat{p})^{R-(i-1)} \hat{p}^{i-1}}{\sum_{i=1}^R\phi_d(i){R\choose R-(i-1)} (1-\hat{p})^{R-(i-1)} \hat{p}^{i-1}},
\label{eqn:relation}
\end{equation}
where \eqnref{eqn:k_i} is inserted cancelling out $\hat{N}$.
An example is
\[
\phi_n(i)=1,
\quad
\phi_d(i)=\left\{\begin{array}{ll}1&\text{if $i=\{2,\ldots,R\}$},\\0&\text{otherwise},\end{array}\right.
\]
which for two reader sessions results in the ratio $(k_1+k_2)/k_2$.

The estimators of $p$ proposed here are defined by their window functions.
As the number of reader sessions increases,
it becomes more likely that elements in $\veck$ becomes zero.
These elements do not provide any information,
and are therefore excluded by setting $\phi_n(i)=\phi_d(i)=0$ when $k_i=0$.
This is used in the numerator window function for both proposed estimators of $p$:
\[
\phi_n(i)=\left\{\begin{array}{ll}1&\text{if $k_i\neq 0$},\\0&\text{otherwise}.\end{array}\right.\\
\]
The difference in the estimators is then the denominator window function.

\subsection*{Estimator 1: Remove Maximum Element}
The first estimator of $p$ is based on the simple principle of removing the largest entry of $\bar{k}$ and all zero elements in the denominator.
This gives the window function:
\[
\phi_d(i)=\left\{\begin{array}{ll}1&\text{if $k_i\neq 0$ and $k_i\neq\max{\veck}$},\\0&\text{otherwise.}\end{array}\right.
\]
This estimator is called the Remove Maximum Element (RME) Estimator.

\subsection*{Estimator 2: Remove Elements Greater than the Mean}
For the second estimator of $p$ an averaged version of $\veck$ is used.
As is shown in \figref{fig:venn},
the two subsets $k_{2a}$ and $k_{2b}$ are added together into $k_2$.
Instead of using this sum, a new vector is defined, $\veck'$, containing estimates of these subsets.
The estimate of the subset is the average of the entries in $\veck$,
with regard to the number of subsets per entry in $\veck$.
This is defined as:
\[
\veck'=\left[\frac{k_1}{{R \choose R}},\frac{k_2}{{R \choose R-1}},\ldots,\frac{k_R}{{R \choose 1}}\right]^T.
\]

The second estimator of $p$ is named the Remove Elements Greater than the Mean (REGM) Estimator.
The denominator window function is
\[
\phi_d(i)=\left\{\begin{array}{ll}1&\text{if $k_i'\neq 0$ and $k_i'<m_{\veck'}$},\\0&\text{otherwise},\end{array}\right.
\]
where $m_{\veck'}$ is the sample mean of the nonzero elements in $\veck'$.
This estimator removes all nonzero elements and all elements greater than $m_{\veck'}$.

\subsection*{The Schnabel Estimator}
We propose to use the simple capture--recapture model,
which provides an estimate of $N$.
When the reader sessions are assumed to be independent,
and as the tags are assumed to be in a closed population,
the tag cardinality estimation can be assumed to be a simple capture--recapture experiment.
When the number of reader sessions, $R$, is two,
the Lincoln-Peterson method provides a maximum likelihood estimate \cite{1991}, where the tag set cardinality is found as
\[
\hat{N}_{LP} = \frac{n_1n_2}{m_2},
\]
where $n_1$ is the number of tags found in the first reader session, $n_2$ is the number of tags found in the second,
and $m_2$ is the number of re-found tags in the second reader session.
For more than two reader sessions, the Schnabel method from \cite{1938} can be used,
which is a weighted average over a series of Lincoln-Peterson estimates
\begin{equation}
\hat{N}_S = \frac{\sum_i^R n_iM_i}{\sum_i^R m_i},
\label{schnabeln}
\end{equation}
where $M_i$ is the total number of tags found in the $(i-1)$th reader session.
Note that the two equations are equal for $R=2$, as $M_1=m_1=0$ and $M_2=n_1$.

The method does not make an intermediate estimate of $p$,
but finds an estimate of $N$ directly.
To compare them and to make an estimate of the error probability $p_M$,
an estimator for $p$, $\hat{p}_S$, is derived.
An estimate of the probability of success for the $i$th reader session is $\frac{n_i}{\hat{N}}$,
and the estimator is found, by averaging over the errors,
\[\hat{p}_S=\frac{1}{R}\sum_{i=1}^R \left(1-\frac{n_i}{\hat{N}_S}\right),\]
which is the sample mean of the error probabilities found in all reader sessions.
This is used for comparison and for calculation of $\hat{p}_M$ as with the other estimators.

\section{Analytical Evaluation}
\label{analytical_analysis}
The analytical work is made for two reader sessions,
as then an explicit estimate of $p$ can be found.
The estimator for $p$ is a function of the observations $k_1$ and $k_2$, denoted $g(k_1,k_2)$:
\begin{equation}
g(k_1,k_2)=\hat{p}=\left\{\begin{array}{ll}1&\text{if $k_1=0$ and $k_2=0$,}\\0&\text{if $k_1>0$ and $k_2=0$,}\\\frac{k_2}{2k_1+k_2}&\text{otherwise,}\\\end{array}\right.
\label{gfunc}
\end{equation}
which follows from \eqnref{ptworeaders},
but with two special cases where either no tags are found or all tags are found in both reader sessions.
Its expected value is given in the following Lemma.
\begin{lem}
\label{explemma}
Let the estimate of $p$ be defined as in \eqnref{gfunc},
then the expected value of $\hat{p}$ for known $N$ and $p$ is
\[
E[g(k_1,k_2)|N,p]=\frac{2N(p-p^{2N})}{2N-1}+p^{2N}.
\]
\end{lem}
\begin{IEEEproof}
See Appendix for the proof.
\end{IEEEproof}
The above result shows that the estimator is biased,
but as $N$ increases and $p$ decreases, then the bias can be neglected.
The bias can in principle be removed,
as it arises due to the definition of the estimator in the marginal cases.
Appropriate choices of the marginal cases can make it unbiased.

The lower limit for $N$ is,
if the expected error made is allowed to be e.g. $1\%$ and $p\leq0.9$,
\[
E[g(k_1,k_2)|N\geq46,p\leq0.9]-p<0.01,
\]
that is, if the maximum assumed error probability is $p=0.9$,
then the minimum number of tags should be $N=46$ to satisfy the error requirement.

The estimate of $N$ is shown to be unbiased in the following.
\begin{lem}
\label{nlemma}
Let the estimate of the tag set cardinality be defined as in \eqnref{ntwocase},
then, for known $N$ and $p$, the estimate of $N$ is unbiased,
that is $E[\hat{N}|N,p]=N$.
\end{lem}
\begin{IEEEproof}
From \eqnref{ntwocase} it follows that
\begin{align*}
E[\hat{N}|N,p]=\sum_{k_1,k_2}\frac{k_1+k_2}{1-p^2}\Pr[K_1=k_1,K_2=k_2],
\end{align*}
and by inserting the multinomial distribution, and the probabilities for each set:
\begin{align*}
E[\hat{N}|N,p]&=\sum_{k_1=0}^N\sum_{k_2=0}^{N-k_1}\frac{k_1+k_2}{1-p^2}{N \choose k_1,k_2,N-k_1-k_2 }\cdot\\
&\qquad(1-p)^{2k_1}(2(1-p)p)^{k_2}p^{2(N-k_1-k_2)}.
\end{align*}
This can be split into two sums, and by the expectation of a multinomial distribution:
\begin{align*}
E[\hat{N}|N,p]&=\frac{1}{1-p^2}\left(E[K_1]+E[K_2]\right)\\
&=\frac{N}{1-p^2}\left((1-p)^2+2(1-p)p\right)=N.
\end{align*}
\end{IEEEproof}
This result ensures,
given a good estimate of the error probability,
that the tag set estimator produces an unbiased tag set cardinality estimate.

For the method to work, the tag sets found in each reader session have to be independent,
as shown in e.g. the examples in Section \ref{introduction}.
To investigate what happens if the reader sessions are dependent,
the estimators are tested in scenarios with dependent reader sessions.
The following section explains how the dependency is modelled,
using a correlation coefficient to specify the correlation between reader sessions.

\subsection{Model for Dependent Reader Sessions}
\label{dependent_reader_sessions}
So far it has been assumed that the reader sessions are independent, but what if this does not hold?
In the following,
a method is introduced to define the correlation for tag $\tau_m$ between the reader sessions $r_i$ and $r_{i+1}$.
For two reader sessions,
define the Bernoulli random variable $X_1$ signifying the outcome of one tag in the first reader session,
and $X_2$ the outcome in the second reader session,
then
\[
X_1=\left\{\begin{array}{ll}1&\text{w.p. $p$,}\\0&\text{w.p. $1-p$,}\\\end{array}\right.
\quad
X_2=\left\{\begin{array}{ll}1&\text{w.p. $pq+(1-p)r$,}\\0&\text{otherwise,}\\\end{array}\right.
\]
where $p$ is the probability of a tag not being read in the first reader session,
$q$ is the probability that it is not read in the second reader session either,
and $r$ is the probability of a tag not being read in the second, but in the first.
This gives the relations:
\begin{align*}
\Pr[X_1=1]&=p & \Pr[X_1=0]&=1-p\\
\Pr[X_2=1|X_1=1]&=q & \Pr[X_2=0|X_1=1]&=1-q\\
\Pr[X_2=1|X_1=0]&=r & \quad\Pr[X_2=0|X_1=0]&=1-r
\end{align*}
It is assumed that the expected error probability remains the same between reader sessions,
because of the random physical displacement of the tags.
Therefore $E[X_1]\equiv E[X_2]$, and
\[pq+(1-p)r=p,\]
where $r$ and $q$ forms the bound $r<p<q$ because an error in the first reader session increases the probability of error in the second.

To specify the level of correlation, the correlation coefficient is used, that is
\[
\rho=\frac{\text{Cov}(X_1,X_2)}{\sigma_{X_1}\sigma_{X_2}}=\frac{q-p}{1-p},
\]
where $0\leq\rho\leq 1$. This yields the correlated probabilities $q$ and $r$ with respect to $p$ and $\rho$ as
\begin{equation}
q=\rho(1-p)+p, \qquad r=\frac{p(1-q)}{1-p}.
\label{corr_qr}
\end{equation}
This is used to show how the presented approach to solve the missing tag problem is affected if the reader sessions are not independent.
The results are shown in the following section.

\section{Simulation Evaluation}
\label{experimental_analysis}
To evaluate the estimators against each other,
and to assert that they perform as expected,
simulations have been carried out.
The true number of tags is set to $N=500$ and each result is averaged over 1000 experiments.

\subsection{Independent Reader Sessions}
\begin{figure}[h]
\centering
\includegraphics[width=8.2cm]{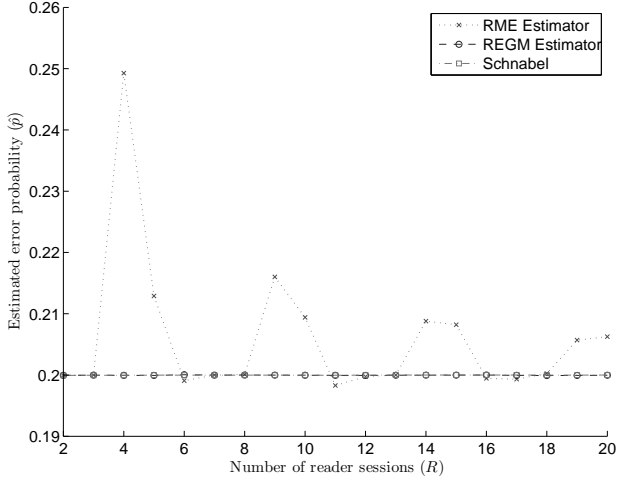}
\caption{Simulated estimation of $p$ vs. number of reader sessions for $p=0.2$.}
\label{fig:three_p}
\end{figure}
The results of the estimate of $p$ are shown in \figref{fig:three_p}.
It shows that the RME Estimator is not performing as good as the others.
This is because the maximum element that is removed may contain almost all the tags and thereby all the information.
By removing it, the estimator makes a bad estimate.
The problem decreases, as the number of reader sessions increase as the tags are spread out in more sets.
Because of the fluctuations for the RME Estimator in its estimate of $p$,
it is not considered further and is not included in any of the following figures.

\begin{figure}[h]
\centering
\includegraphics[width=8.2cm]{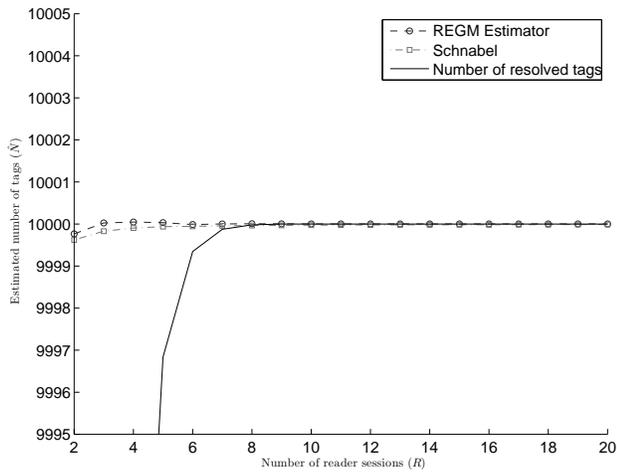}
\caption{Simulated estimation of $N$ vs. number of reader sessions for $p=0.2$.}
\label{fig:three_n}
\end{figure}
\begin{figure}[h]
\centering
\includegraphics[width=8.2cm]{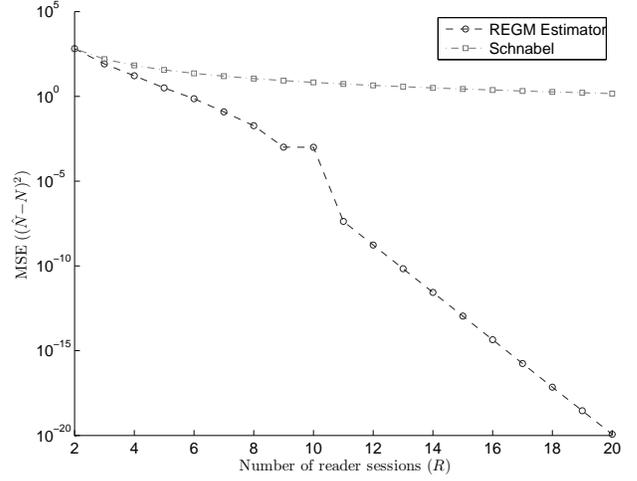}
\caption{Simulated MSE of $N$ vs. number of reader sessions for $p=0.2$.}
\label{fig:two_mse}
\end{figure}
The tag set cardinality is estimated in \figref{fig:three_n}.
The estimate given by the two estimators is similar, but the REGM Estimator converges faster to the true number of tags.
This can be seen in \figref{fig:two_mse},
where the mean-square error of $N$ is given,
showing that the Schnabel Estimator converges to zero more slowly than the REGM Estimator.

\begin{figure}[h]
\centering
\includegraphics[width=8.2cm]{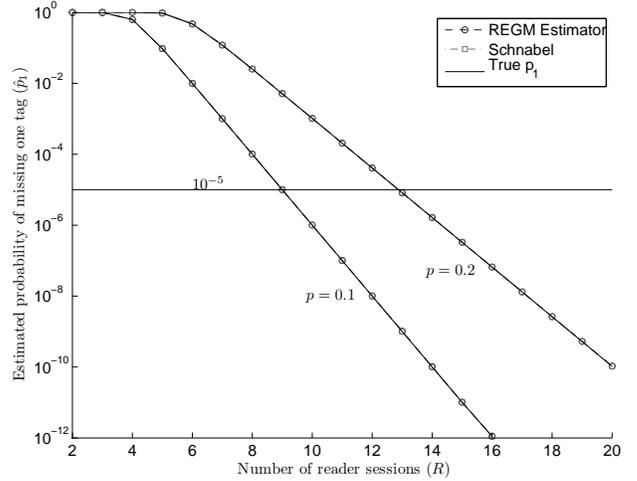}
\caption{Simulated estimation of $p_M$ vs. number of reader sessions for $p=0.1$ and $p=0.2$.
An example threshold is at $10^{-5}$.}
\label{fig:two_term}
\end{figure}
The estimate of $p_M$ is the most important estimate,
as it shows how many reader sessions are needed to be certain,
with high probability, that all tags are resolved.
Results are in \figref{fig:two_term} for $p=0.1$ and $p=0.2$.
It can be seen that both estimators are close to the true $p_M$ calculated using \eqnref{eqn:p_m} using true $p$ and $N$ as if they were known a priori.
Therefore, if the error probability is $p=0.1$,
then the sequential decision process determines to stop after $R=8$ reader sessions,
and for $p=0.2$ it is $R=12$, if the allowed threshold is $10^{-5}$.
The $p=0.2$ case can be compared with \figref{fig:three_n},
where it is seen, that all tags are found in approximately $8$ reader sessions.

\subsection{Dependent Reader Sessions}
In the following the estimators are tested in scenarios where the independence assumption does not hold.
For the simulations it is chosen to use $\rho=0.1$ and $\rho=0.3$,
to demonstrate the effect of correlated reader sessions.
The correlated error probabilities are found using \eqnref{corr_qr},
in which the correlation coefficient $\rho$ is a parameter.

\begin{figure}[h]
\centering
\includegraphics[width=8.2cm]{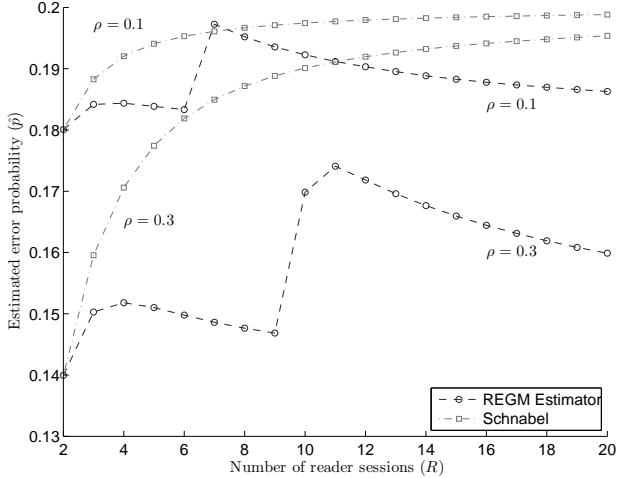}
\caption{Simulated estimation of $p$ vs. number of reader sessions with correlation between the reader sessions. $p=0.2$ and the correlation coefficient is $\rho=0.1$ and $\rho=0.3$.}
\label{fig:p_corr}
\end{figure}
The estimated error probabilities are shown in \figref{fig:p_corr},
where it can be seen, that the estimators are affected by the correlated reader sessions.
The Schnabel Estimator converges to the correct error probability,
where the REGM Estimator converges to some other error probability,
depending on the correlation.

\begin{figure}[h]
\centering
\includegraphics[width=8.2cm]{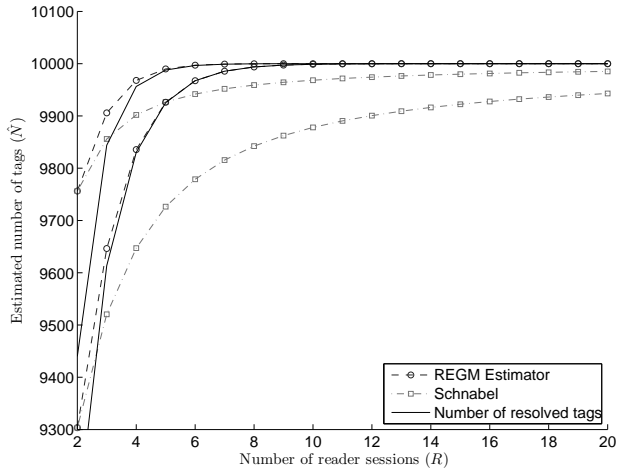}
\caption{Simulated estimation of $N$ vs. number of reader sessions with correlation between the reader sessions. $p=0.2$ and the correlation coefficient is $\rho=0.1$ for the upper values, and $\rho=0.3$ for the lower values.}
\label{fig:n_corr}
\end{figure}
Even though the error probability estimates for the REGM Estimator converges to wrong values of $\hat{p}$,
it performs better than the Schnabel Estimator when estimating the tag set cardinality.
This is seen in \figref{fig:n_corr},
where the REGM Estimator never provides an estimate lower then the actual number of resolved tags.
Both estimators converges slower towards the true $N$, because of the correlation between the reader sessions.

\begin{figure}[h]
\centering
\includegraphics[width=8.2cm]{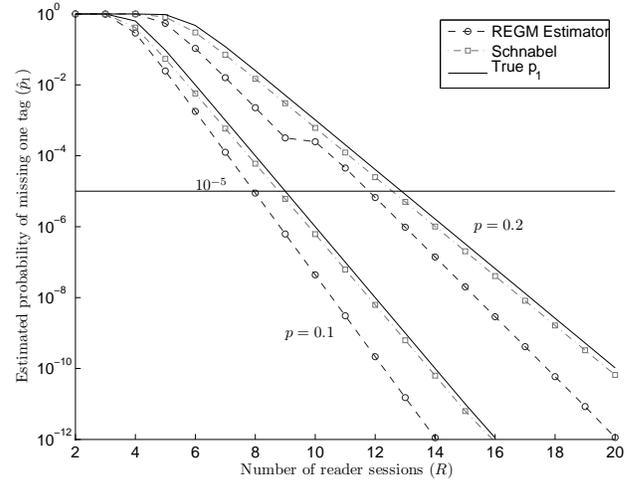}
\caption{Simulated estimation of $p_M$ vs. number of reader sessions with correlation between the reader sessions. $p=0.1$ and $p=0.2$, and the correlation coefficient is $\rho=0.3$. An example threshold is at $10^{-5}$.}
\label{fig:p_M_corr}
\end{figure}
The estimate of the probability of completely missing one tag is shown in \figref{fig:p_M_corr},
where it can be seen, that the correlation affects the performance of the estimate.
Therefore if the estimator is used as is, the estimate is wrong.
The ideas for a solution to this, proposed in Section \ref{framework},
is to make some estimation margin, e.g. two additional reader sessions,
so that more reader sessions than strictly necessary is used,
to be certain the probability of missing one or more tags is below the chosen threshold.

\section{Conclusion}
\label{conclusion}
In this paper two different methods for obtaining the error probability estimate and the tag set cardinality estimate are proposed.
The first method, named the REGM Estimator, is based on the assumption that it is possible to obtain statistically independent, uncorrelated reader sessions.
First this estimator is introduced and explained with two reader sessions,
after which it is extended to the general case.
Then a method is devised to calculate the number of required reader sessions to guarantee,
with some probability, that no tags are missing.
The second estimator is based on the Schnabel method, known from capture--recapture literature,
which is extended to also provide estimates of the error probability and the probability that tags are missing.

It is shown that the REGM Estimator for the error probability for two reader sessions is biased,
but that the bias becomes insignificant when the number of tags increases and the error probability decreases.
Also, it is shown that the estimate of the tag set cardinality is unbiased in the case of two reader sessions.
For the estimators to work it is important that the assumption of independent, uncorrelated reader sessions holds.
To show how the estimators behave when the reader sessions are correlated, a model is devised for use in the simulations.

Simulations are performed,
which show that the tag set cardinality estimator using the estimated error probability from the REGM Estimator converges towards the correct value faster than the Schnabel Estimator.
They also show that more reader sessions decreases the probability of a missing tag,
indicating that the proposed method for estimating the probability of missing a tag is working.
Experiments with dependent reader sessions show that the estimation of the tag set cardinality requires more reader sessions to be precise,
but that the REGM Estimator's estimate of the tag set cardinality still converges faster than the one based on the Schnabel method.
However both estimators underestimate the probability of missing one or more tags,
resulting in a possibility of premature termination of the algorithm.
To counter this, some estimation margin should be used when the reader sessions are dependent,
and the analysis of this margin will be investigated in further work.
Another interesting venue for future work is to investigate the cases when the reading errors have correlations across the tags in the same reader session. The future work should include evaluation of the proposed methods by using more detailed physical models for the tag reading errors.

\bibliographystyle{IEEEtran}
\bibliography{bibtex}

\begin{appendix}
The following is the proof of Lemma \ref{explemma}.
\begin{IEEEproof}
The estimator is defined as in \eqnref{gfunc} and the expected value $E[g(K_1,K_2)|N,p]$ is
\begin{align*}
E[g(K_1,K_2)]&=\sum_{k_1,k_2}g(k_1,k_2)\Pr[K_1=k_1,K_2=k_2]\\
&=\sum_{k_1=0}^N\sum_{k_2=1}^{N-k_1}\frac{k_2}{2k_1+k_2}\Pr[K_1=k_1,K_2=k_2]+\\
&\qquad\Pr[K_1=0,K_2=0].
\end{align*}
We insert the multinomial distribution with the probabilities for each set,
\begin{align*}
E[g(K_1,K_2)]&=\sum_{k_1=0}^N\sum_{k_2=1}^{N-k_1}\frac{k_2}{2k_1+k_2}{ N \choose k_1,k_2,N-k_1-k_2}\cdot\\
&\qquad(1-p)^{2k_1}(2(1-p)p)^{k_2}p^{2(N-k_1-k_2)}+p^{2N}\\
&=p^{2N}\sum_{k_1=0}^N\sum_{k_2=1}^{N-k_1}2^{k_2}\frac{k_2}{2k_1+k_2}\cdot\\
&\qquad{ N \choose k_1,k_2,N-k_1-k_2}\left(\frac{1-p}{p}\right)^{2k_1+k_2} + p^{2N}.
\end{align*}
We define a function $h$, which is all but the two $p^{2N}$,
and we differentiate it with respect to $p$,
\begin{align*}
\frac{dh}{dp}&=\sum_{k_1=0}^N\sum_{k_2=0}^{N-k_1}2^{k_2}k_2{ N \choose k_1,k_2,N-k_1-k_2}\cdot\\
&\qquad\left(\frac{1-p}{p}\right)^{2k_1+k_2-1}\left(\frac{-1}{p^2}\right)\\
&=-\frac{1}{(1-p)p^{2N+1}}\cdot\\
&\qquad\sum_{k_1=0}^N\sum_{k_2=0}^{N-k_1}k_2{ N \choose k_1,k_2,N-k_1-k_2}\cdot\\
&\qquad(1-p)^{2k_1}(2(1-p)p)^{k_2}p^{2(N-k_1-k_2)}\\
&=-\frac{1}{(1-p)p^{2N+1}}E[K_2]\\
&=-\frac{1}{(1-p)p^{2N+1}}N2(1-p)p=-\frac{N2}{p^{2N}}.
\end{align*}
This function is integrated and merged with the part not differentiated, this gives
\begin{align*}
h&=\int-\frac{2N}{p^{2N}}\text{d}p=\frac{2Np^{1-2N}}{2N-1}+c\\
E[g(K_1,K_2)]&=p^{2N}\left(\frac{2Np^{1-2N}}{2N-1}+c\right) + p^{2N}.
\end{align*}
By inserting known $p$, and solving with respect to the integral coefficient $c$, $c$ is found to $-\frac{2N}{2N-1}$,
and the expected value is
\begin{align*}
E[g(K_1,K_2)]&=p^{2N}\left(\frac{2N(p^{1-2N}-1)}{2N-1}\right)+p^{2N}\\
&=\frac{2N(p-p^{2N})}{2N-1} + p^{2N},
\end{align*}
which is approximately $p$ for large $N$.
\end{IEEEproof}
\end{appendix}

\end{document}

%% file: venn.tex
\ifx\JPicScale\undefined\def\JPicScale{1}\fi
\unitlength \JPicScale mm
\begin{picture}(50,35)(0,0)
\put(25,0){\makebox(0,0)[cc]{$k_2$}}

\put(15,20){\circle{30}}
\put(35,20){\circle{30}}
\put(20,8){\line(1,-1){5}}
\put(30,8){\line(-1,-1){5}}

\put(15,18){\hskip-1.9mm\includegraphics[scale=0.1]{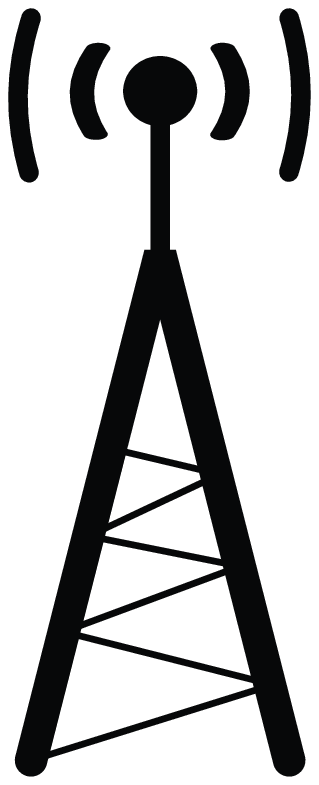}}
\put(15,16){\makebox(0,0)[cc]{$\mathbf{r_1}$}}
\put(35,18){\hskip-1.9mm\includegraphics[scale=0.1]{antenna}}
\put(35,16){\makebox(0,0)[cc]{$\mathbf{r_2}$}}

\put(5,20){\makebox(0,0)[cc]{$k_{2a}$}}
\put(45,20){\makebox(0,0)[cc]{$k_{2b}$}}

\put(25,20){\makebox(0,0)[cc]{$k_1$}}
\put(25,35){\makebox(0,0)[cc]{$k_3$}}

\put(45,10){\makebox(0,0)[cc]{}}

\end{picture}